\documentclass[baaa]{baaa}

\usepackage[pdftex]{hyperref}
\usepackage{subfigure}
\usepackage{natbib}
\usepackage{helvet,soul}
\usepackage[font=small]{caption}

\begin{document}


\journalvol{61A}
\journalyear{2019}
\journaleditors{R. Gamen, N. Padilla, C. Parisi, F. Iglesias \& M. Sgr\'o}


\contriblanguage{1}


\contribtype{1}

\thematicarea{7}

\title{Hybrid magnetized stars within the Field Correlator Method}


\titlerunning{Hybrid magnetized stars within the FCM}


\author{Mariani M.\inst{1,3}, Orsaria M.G.\inst{1,3}, Ranea-Sandoval I.F.\inst{1,3} \& Guilera O.M.\inst{2,4,5}}
\authorrunning{Mariani et al.}


\contact{mmariani@fcaglp.unlp.edu.ar}

\institute{Grupo de Gravitaci\'on, Astrof\'isica y Cosmolog\'ia,  Facultad de Ciencias Astron\'omicas y Geof\'isicas, \\ Universidad Nacional de La Plata, Paseo del Bosque S/N (1900), La Plata, Argentina.\and Facultad de Ciencias Astron\'omicas y Geof\'isicas, \\ Universidad Nacional de La Plata, Paseo del Bosque S/N (1900), La Plata, Argentina. \and CONICET, Godoy Cruz 2290, 1425,  Buenos Aires, Argentina. \and Instituto de Astrof\'isica de La Plata, CONICET, Argentina. \and Instituto de Astrof\'isica, Pontificia Universidad Católica de Chile, Santiago, Chile.}


\resumen{
Presentamos resultados preliminares del estudio de los efectos de campos magnéticos intensos en estrellas híbridas. Para la descripción de la fase hadrónica, utilizamos la aproximación de campo medio relativista. Para la materia de quarks, empleamos el formalismo del Field Correlator Method. Construida la ecuación de estado, discutimos la anisotropía de las presiones debido a la presencia del campo magnético. Finalmente, calculamos la estructura de las estrellas compactas utilizando las ecuaciones de estado híbridas magnetizadas y sus modos de oscilación relacionados con la emisión de ondas gravitacionales.}

\abstract{
We present preliminary results of the study of intense magnetic fields effects on hybrid stars. For the description of the hadronic phase, we use the relativistic mean field approximation. For the quark matter phase, we employ the Field Correlator Method formalism. Once the the equation of state is built, we discuss the pressure anisotropy due the presence of the strong magnetic field. Finally, we calculate the structure of the compact stars using magnetized hybrid equations of state and their oscillation modes related with the emission of gravitational waves.
}



\maketitle

\section{Introduction}
\label{S_intro}

    
Neutron stars (NSs) or hybrid stars (HSs) are extremely compact ($M \sim 1.4$~M$_\odot$, $R \sim 10$~km) remnants of high-mass stars. In their inner cores, matter is compressed to densities larger than $10^{15}$~g/cm${}^3$ so new exotic particles like mesons, hyperons and/or deconfined quarks could appear. The discovery of the $2$~M$_\odot$ pulsars PSR J1614-2230 \cite{demo} and PSR J0348-0432 \cite{anto} has imposed restrictions to the equation of state (EoS) needed to describe matter inside compact objects and has forced astrophysics to rethink the internal composition of NSs.
    
    
On the other hand, it is well known that at the surface of NSs there exist magnetic fields (MFs) of the order of $10^{11-13}$~Gauss (the “classical pulsars”). However, there is evidence of NSs with ultra strong MFs, called ``magnetars'', in which the surface MF could be higher than $10^{14-15}$~G.


In addition, the recent direct detections of gravitational waves (GWs) have made GW astronomy become highly relevant. In this way, it is known that the NS oscillation modes are directly related to its interior composition \citep{cow1}. In particular, non-radial oscillation modes may emit GWs. For this reason, the study of non-radial pulsations and their associated GW emission could shed some light on the behavior of matter inside NSs.


\section{Hybrid EoS and phase transition}

We model cold HSs using a hybrid-magnetized EOS by combining the magnetized BPS EoS for the crust \citep{crust}, the magnetic non-linear relativistic mean field model \citep{rmf} for the outer hadronic core and a magnetized version of the Field Correlator Method (FCM) EoS of quark matter for the inner core \citep{fcm, quamag}. We consider a chemical potential, $\mu_b$, dependent magnetic field in the $z$-direction given by \citep{camposime}:
\begin{equation}
	B(\mu_b) = B_{\text{surface}}+B_{\text{center}}[1-e^{\beta\frac{(\mu_b-m_n)^\alpha}{m_n}}] \,
\end{equation}
where $\alpha=2.5$, $\beta = -4.08 \ 10^{-4}$ are fixed parameters and $m_n$ is the nucleon mass.


The FCM quark model is characterized by two parameters: the gluon condensate, $G_2$, and the large distance static $\bar{q}q$ potential, $V_1$.

As we are studying cold hybrid stars, the only leptons present in neutron star matter in chemical equilibrium are electrons and muons. 
    

To build the phase transition between hadronic and quark matter, we use the Maxwell construction in which no mixed phase exists and there is a sharp discontinuity. Electric charge is conserved locally and $\beta$-equilibrium condition, which establishes equilibrium among weak interaction processes, is also taken into account.

We study two different MF configurations (consistent with a ``classical pulsar'', A1 and B1 cases, and with a ``magnetar'', A2 and B2 cases) and two sets of the FCM parameters. Details of the parameter values for each case are given in Table \ref{tabla1}.

\begin{figure}[ht]
\centering
\includegraphics[width=0.87\columnwidth]{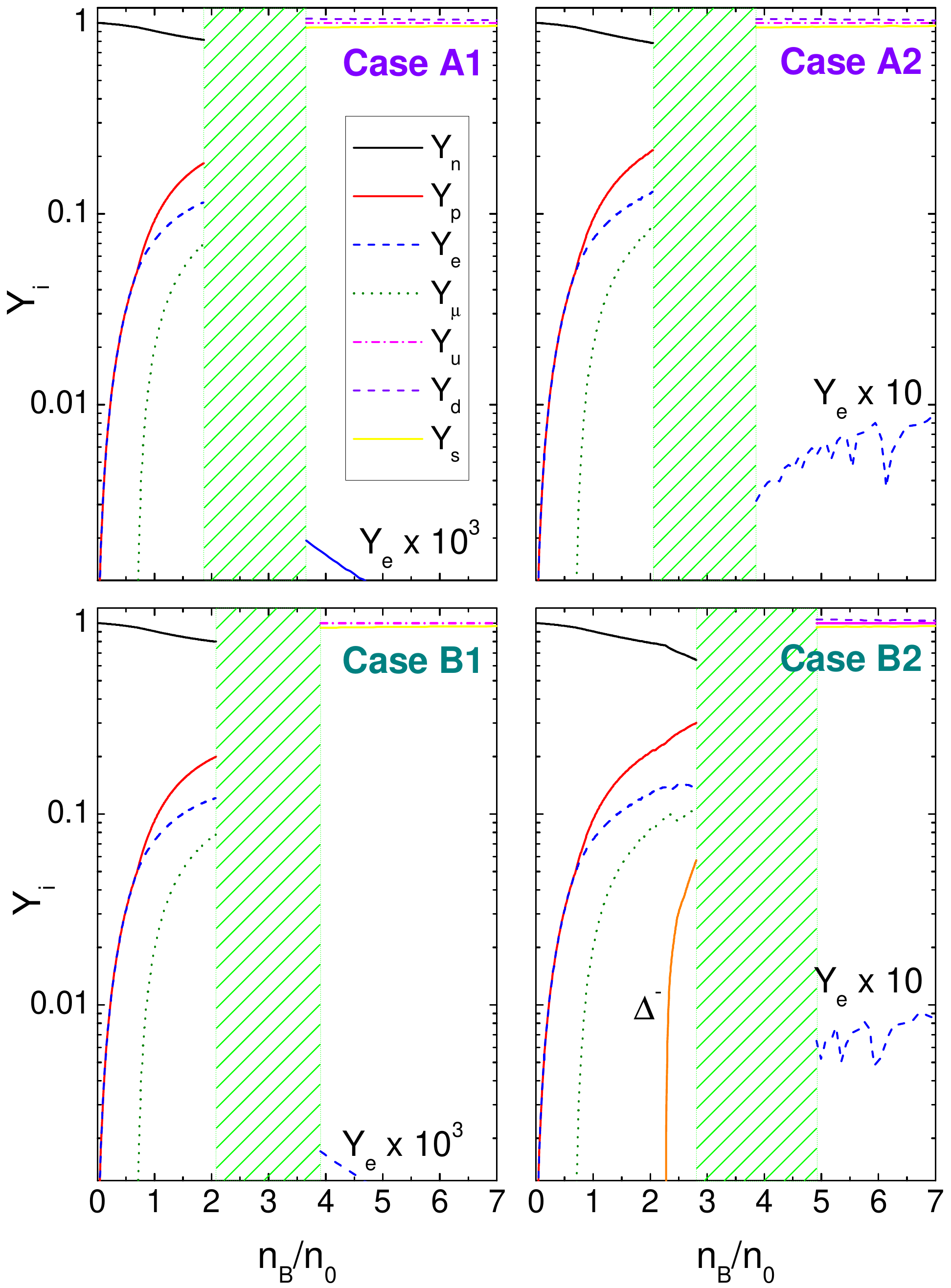}
\caption{Particle populations as a function of baryon number density in nuclear density units ($n_0 = 0.16$~fm$^{-3}$). The left region of each panel corresponds to the hadronic phase and the right one shows the quark phase population. The green area represents the jump in the density corresponding to the phase transition.}
\label{popu}
\end{figure}

Considering these conditions, it is possible to construct hybrid EoS which determine the particle population of magnetized HSs, as shown in Fig. \ref{popu}. It can be seen that the MF shifts the phase transition to high densities. This is more noticeable by comparing B1 and B2 cases in which a $\Delta^{-}$ appears in the hadronic phase as the phase transition to quark matter occurs later. The strong MF also increases the electron population in the quark phase.


One of the strong MF effects is to produce an anisotropy on the pressure, related to its axial direction \citep{presion1, presion2}:
\begin{equation}
	P_\parallel= P_{matter} - \epsilon_{matter} - \frac{B^2}{2} \, ,
\end{equation}
\begin{equation}
	P_\perp= P_{matter} - \epsilon_{matter} + \frac{B^2}{2} - \mathcal{M}B \, ,
\end{equation}
where $\mathcal{M}=-\frac{\partial \Omega}{\partial B}$ ($\Omega$ is the Grand canonical potential) is the matter magnetization and $\epsilon_{matter}$ is the matter energy density. As we will explain in detail later, we have considered an average of the pressure tensor components to obtain the total pressure of the system.

\begin{table}[!h]
\centering
\caption{Parameters values and labels for each studied case.}
\begin{tabular}{lcccc}
\hline\hline\noalign{\smallskip}
Case & $B_{\text{surface}}$ & $B_{\text{center}}$ & $V_1$ & $G_2$ \\
& $[\text{Gauss}]$ & $[\text{Gauss}]$ & $[\text{MeV}]$ & $[\text{GeV}^4]$ \\
\hline\noalign{\smallskip}
A1 & $1.0 \ 10^{13}$ & $1.0 \ 10^{15}$ & $15$ & $0.006$ \\ 
A2 & $1.0 \ 10^{15}$ & $5.0 \ 10^{18}$ & $15$ & $0.006$ \\
B1 & $1.0 \ 10^{13}$ & $1.0 \ 10^{15}$ & $20$ & $0.006$ \\
B2 & $1.0 \ 10^{15}$ & $5.0 \ 10^{18}$ & $20$ & $0.006$	\\
\hline
\end{tabular}
\label{tabla1}
\end{table}

In the top panel of Fig. \ref{eosmraio}, we show the resulting hybrid EoS for the four cases considered.




\section{Hybrid star structure}

Once the hybrid EoS is obtained, we construct the family of stationary stellar configurations. The Tolman-Oppenheimer-Volkoff (TOV) equations are the relativistic structure equations of hydrostatic equilibrium and mass conservation for a spherically symmetric space-time. These equations determine the mass-radius relationship for each family of stars.

\begin{figure}[h]
  \centering
  \includegraphics[width=0.8\columnwidth]{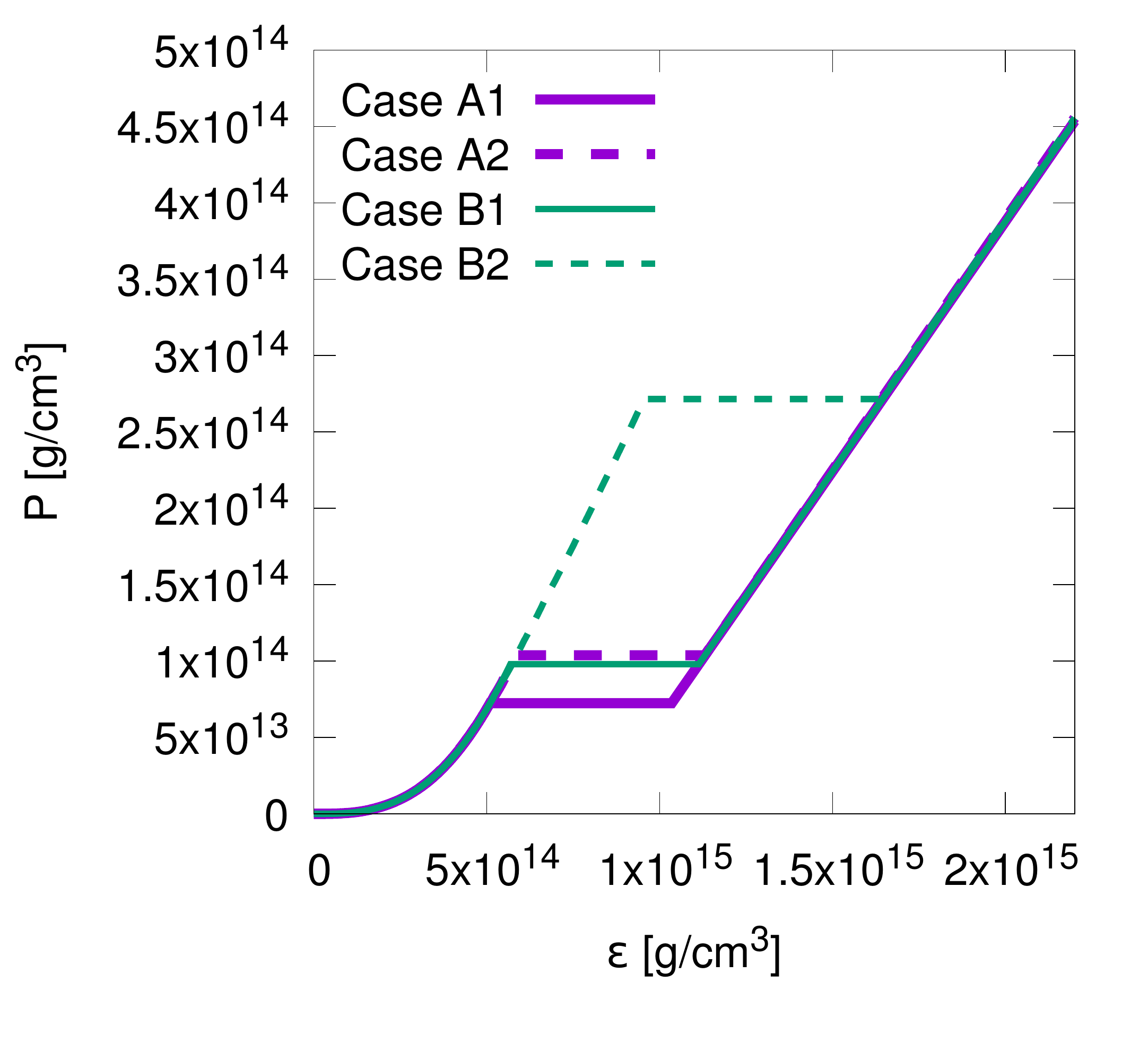}
  \includegraphics[width=0.8\columnwidth]{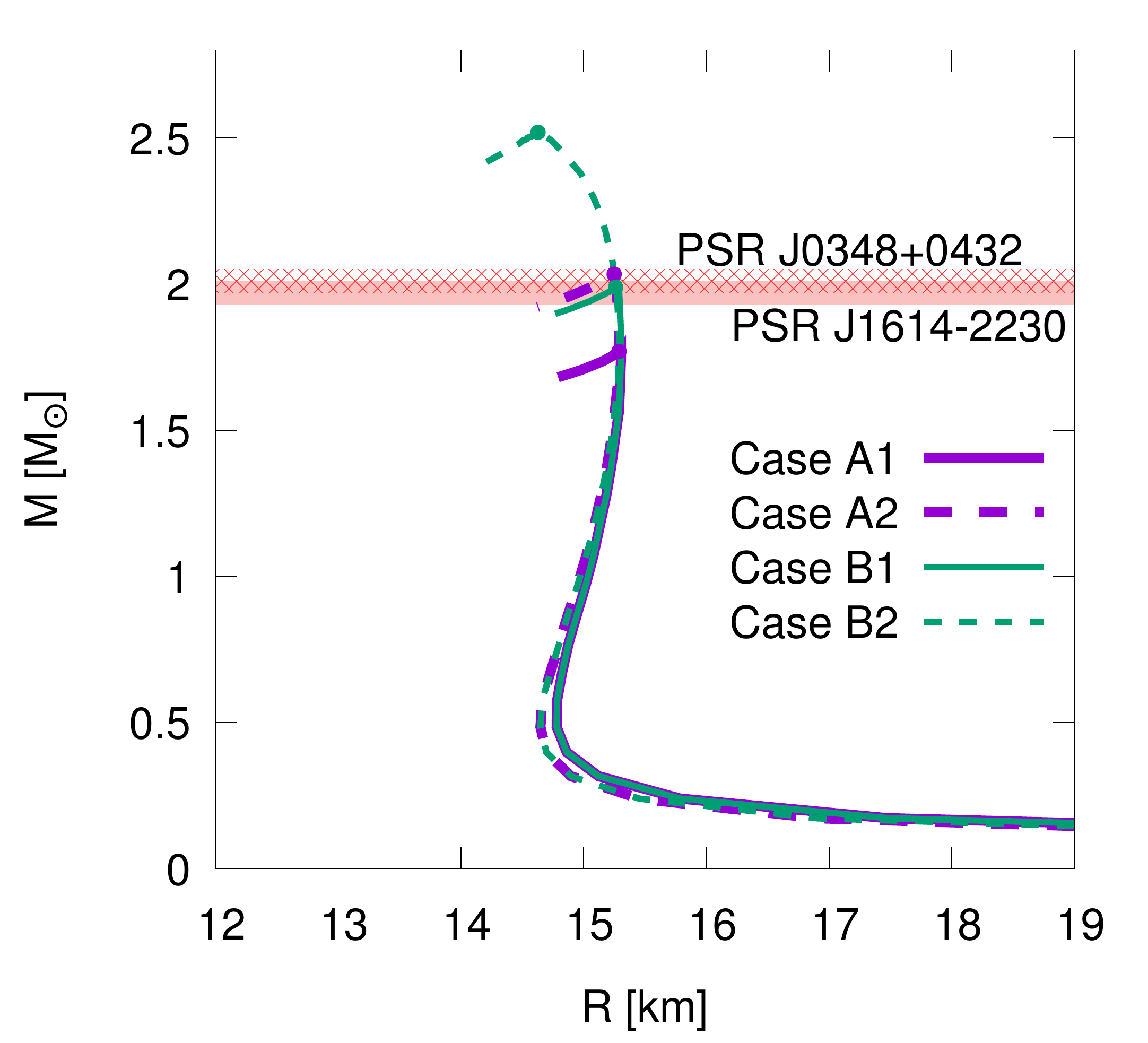}
  \caption{Hybrid EoSes (top panel) and M-R relationship (bottom) for the four cases. In the EoS curves, the constant pressure regions correspond to the transition phase. It can be seen that the effect of the strong MF delays the phase  transition and stiffens the EoSes. In the M-R curves, the rounded dot indicates where the quark mater core appears. After the peaks, towards smaller radii, the stars become unstable. The horizontal bars are the measured masses of the $2$~M$_\odot$ pulsars with their corresponding errors. It can be seen that the magnetars reach higher maximum masses than the classical pulsars.}
  \label{eosmraio}%
\end{figure}

In our case, the MF breaks the spherical symmetry. Including the breaking of the spherical symmetry in a consistent way would lead to 2D calculations. To avoid this complication, we have considered the total pressure as the average of the pressure contributions in different directions, maintaining in this way the isotropy of the system \citep{camposime}. With this choice, we are able to use TOV equations without any change and study the effect on the M-R relationships for magnetized HS. We present our results in Fig.\ref{eosmraio}.

\begin{figure}[h]
 \centering
 \includegraphics[width=0.7\columnwidth, angle=-90]{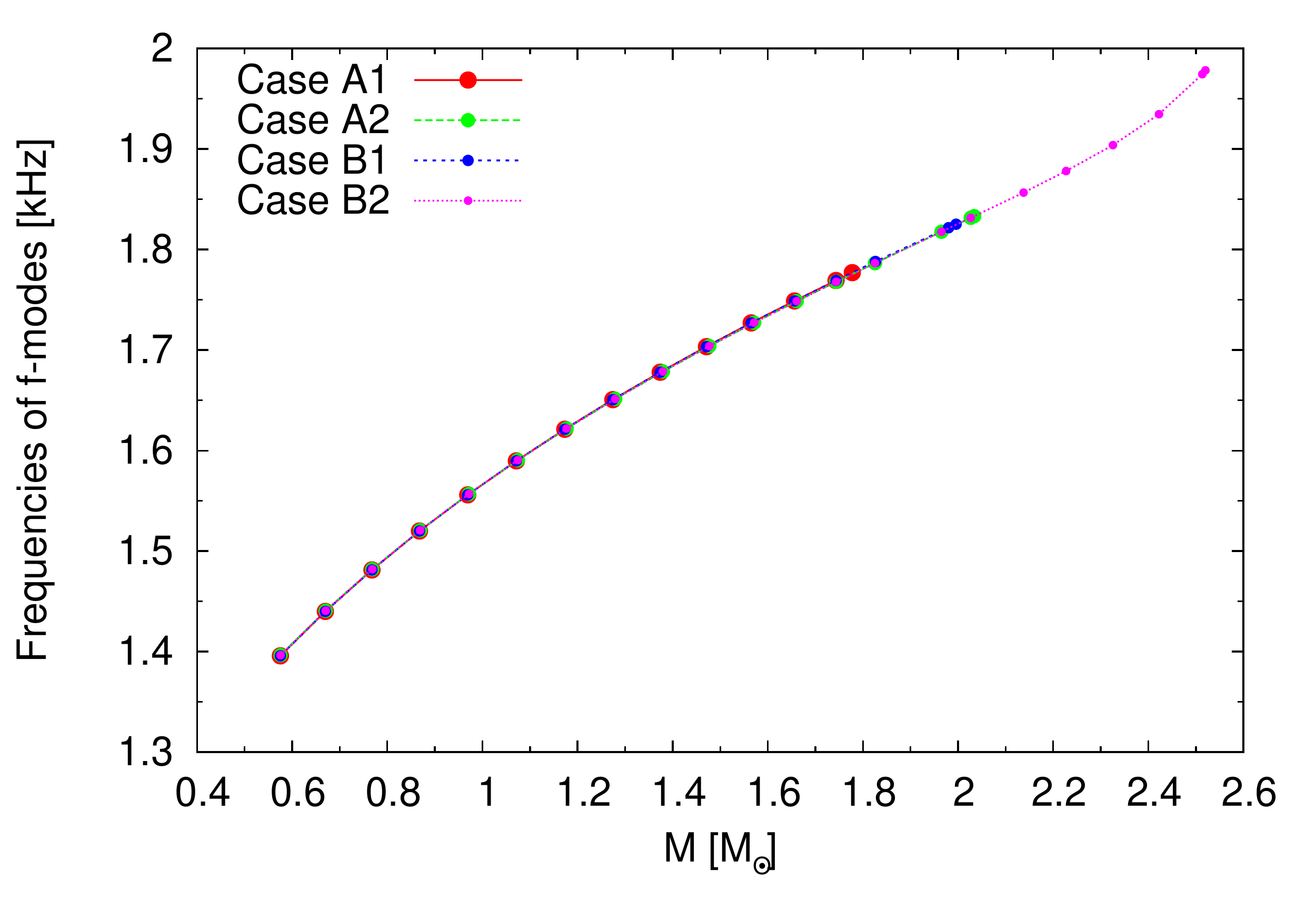}
 \includegraphics[width=0.7\columnwidth, angle=-90]{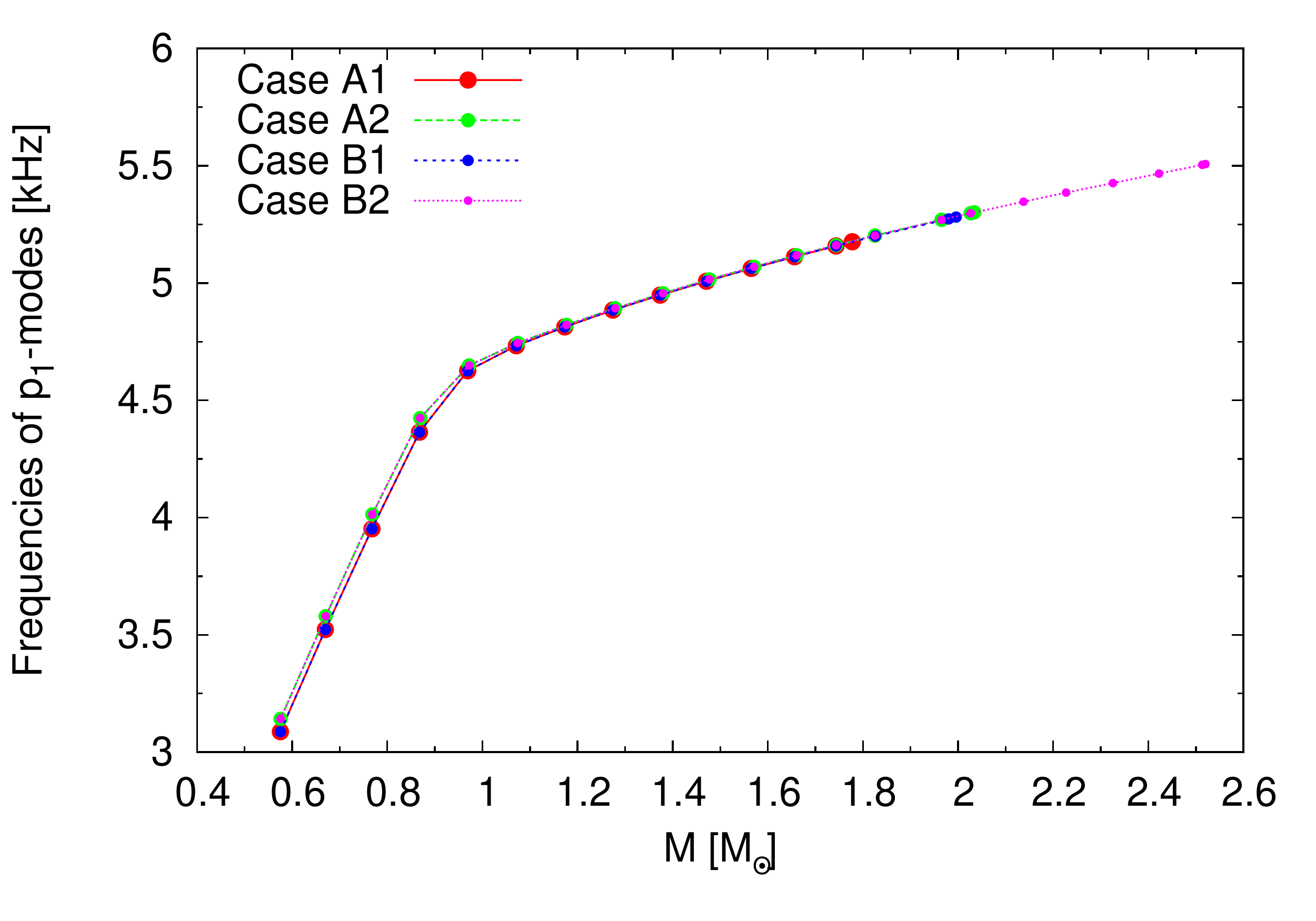}
 \caption{Frequencies of $f$ (top panel) and $p_1$ (bottom) modes for the four cases. The results are almost coincident due to the low MF along the hadronic phase.}
 \label{modesfp}%
\end{figure}

In addition, we study stellar polar pulsations within the relativistic Cowling approximation \citep{cow1, cow2, cownos} where metric perturbations are neglected. In Fig. \ref{modesfp} we present the frequencies of the fundamental ($f$) and first pressure ($p_1$) modes as a function of the stellar mass. The four cases show non distinguishable frequencies. As the frequency calculations are possible only for the -purely hadronic- stable stars, because the appearance of quark matter immediately destabilizes the star, and as the MF increases for higher densities, these results are consequence of the negligible effect of the low MF in the hadronic phase.

\section{Conclusions and discussion}
     
We have calculated magnetic hybrid EoSes and we have obtained stable magnetars that reach $2$~M$_\odot$ for the maximum mass star. According to these results, the magnetic field would increase the maximum mass star of the stable branch of the HSs family in comparison to the classical pulsars. This is a consequence of a late phase transition from hadronic to quark matter in the magnetized neutron star matter.

Regarding the FCM parameters, the increase of $V_1$ produces an increase in the maximum mass of the last stable star. Remains to analyze the effect of a $G_2$ variation.
     
For the four cases studied, the appearance of the quark matter destabilizes the star. In this way, all the stable stars obtained are pure hadronic stars.

From the analysis of the oscillation modes, as we have explained before, it turns out that all the cases studied show very similar frequencies values. These results suggest that the detection of $f$ and $p_1$ modes would not be enough to distinguish between classical pulsars or magnetars. Nevertheless, as these are a preliminary results, it is necessary to study new values for the model parametrizations, consider the existence of a mixed hadron-quark phase and calculate other frequency oscillation modes, as the g-mode.

\begin{acknowledgement}
MM, MGO and IFRS acknowledge support from Universidad Nacional de La Plata and CONICET under Grants G140, G157 and PIP-0714. OMG thanks ANPCyT and UNLP for financial support under Grants PICT 2006-0053 and G157.
\end{acknowledgement}


\bibliographystyle{baaa}
\small
\bibliography{biblio_mariani}
 
\end{document}